\documentclass[aps,prl,twocolumn,superscriptaddress,twoside, nofootinbib]{revtex4-1}
\usepackage{amsmath,amssymb,amsfonts}
\usepackage{graphicx}
\usepackage[english]{babel}
\usepackage{psfrag}
\usepackage{color}
\usepackage{tabularx}
\usepackage{times}
\usepackage{hyperref}
\usepackage{float}

\begin{document}

\title{Quantum phases of SrCu$_2$(BO$_3$)$_2$ from high-pressure thermodynamics}

\author{Jing Guo}
\affiliation{Beijing National Laboratory for Condensed Matter Physics and Institute of Physics, Chinese Academy of Sciences, Beijing 100190, China}

\author{Guangyu Sun}
\affiliation{Beijing National Laboratory for Condensed Matter Physics and Institute of Physics, Chinese Academy of Sciences, Beijing 100190, China}
\affiliation{School of Physical Sciences, University of Chinese Academy of Sciences, Beijing 100190, China}

\author{Bowen Zhao}
\affiliation{Department of Physics, Boston University, 590 Commonwealth Avenue, Boston, Massachusetts 02215, USA}

\author{Ling Wang}
\affiliation{Beijing Computational Science Research Center, 10 East Xibeiwang Road, Beijing 100193, China}
\affiliation{Zhejiang Institute of Modern Physics, Zhejiang University, Hangzhou 310027, China}

\author{Wenshan Hong}
\affiliation{Beijing National Laboratory for Condensed Matter Physics and Institute of Physics, Chinese Academy of Sciences, Beijing 100190, China}
\affiliation{School of Physical Sciences, University of Chinese Academy of Sciences, Beijing 100190, China}

\author{Vladimir A. Sidorov}
\affiliation{Vereshchagin Institute for High Pressure Physics, Russian Academy of Sciences, 108840 Troitsk, Moscow, Russia}

\author{Nvsen Ma}
\affiliation{Beijing National Laboratory for Condensed Matter Physics and Institute of Physics, Chinese Academy of Sciences, Beijing 100190, China}

\author{Qi Wu}
\affiliation{Beijing National Laboratory for Condensed Matter Physics and Institute of Physics, Chinese Academy of Sciences, Beijing 100190, China}

\author{Shiliang Li}
\affiliation{Beijing National Laboratory for Condensed Matter Physics and Institute of Physics, Chinese Academy of Sciences, Beijing 100190, China}
\affiliation{School of Physical Sciences, University of Chinese Academy of Sciences, Beijing 100190, China}
\affiliation{Songshan Lake Materials Laboratory, Dongguan, Guangdong 523808, China}

\author{Zi Yang Meng}
\email{zymeng@iphy.ac.cn}
\affiliation{Beijing National Laboratory for Condensed Matter Physics and Institute of Physics, Chinese Academy of Sciences, Beijing 100190, China}
\affiliation{Department of Physics and HKU-UCAS Joint Institute of Theoretical and Computational Physics, The University of Hong Kong, Pokfulam Road, Hong
Kong, China}
\affiliation{Songshan Lake Materials Laboratory, Dongguan, Guangdong 523808, China}

\author{Anders W. Sandvik}
\email{sandvik@bu.edu}
\affiliation{Department of Physics, Boston University, 590 Commonwealth Avenue, Boston, Massachusetts 02215, USA}
\affiliation{Beijing National Laboratory for Condensed Matter Physics and Institute of Physics, Chinese Academy of Sciences, Beijing 100190, China}

\author{Liling Sun}
\email{llsun@iphy.ac.cn}
\affiliation{Beijing National Laboratory for Condensed Matter Physics and Institute of Physics, Chinese Academy of Sciences, Beijing 100190, China}
\affiliation{School of Physical Sciences, University of Chinese Academy of Sciences, Beijing 100190, China}
\affiliation{Songshan Lake Materials Laboratory, Dongguan, Guangdong 523808, China}

\date{\today}

\begin{abstract}
We report heat capacity measurements of SrCu$_2$(BO$_3$)$_2$ under high pressure along with 
simulations of relevant quantum spin models and map out the $(P,T)$ phase diagram of the material. 
We find a first-order quantum phase transition between the low-pressure quantum dimer paramagnet 
and a phase with signatures of a plaquette-singlet state below T = 2 K. At higher pressures,
we observe a transition into a previously unknown antiferromagnetic state below 4 K. Our findings 
can be explained within the two-dimensional Shastry-Sutherland quantum spin model supplemented by 
weak inter-layer couplings. The possibility to tune SrCu$_2$(BO$_3$)$_2$ between the plaquette-singlet 
and antiferromagnetic states opens opportunities for experimental tests of quantum field theories 
and lattice models involving fractionalized excitations, emergent symmetries, and gauge 
fluctuations.
\end{abstract}

\maketitle

Theoretical proposals for exotic states in quantum magnets abound \cite{wen19,senthil04,sachdev08,shao16,Zhao18,senthil17}, but many intriguing quantum 
phases and transitions beyond classical descriptions have been difficult to realize experimentally. In one class of hypothetical states, spins 
entangle locally and form symmetry-breaking singlet patterns \cite{senthil17,senthil04,sachdev08,shao16,haldane88,read89,capriotti00,koga00,Zhao18}.
Signatures of a state with four-spin singlets were recently detected in the two-dimensional (2D) quantum magnet SrCu$_2$(BO$_3$)$_2$ 
under high pressure \cite{Zayed2017}. This plaquette singlet (PS) state has remained controversial, however \cite{Boos2019}, and a putative phase 
transition into an antiferromagnet (AF) at still higher pressure has not been studied. In this Letter, we report the phase diagram of SrCu$_2$(BO$_3$)$_2$ 
based on heat capacity measurements for a wide range of pressures $P$ and temperatures $T$ down to $0.4$ K. Copmparing the results with calculations
for relevant quantum spin models, our results indicate a PS--AF transition between $P=2.5$ and $3$ GPa, which is significantly lower than 
previously anticipated \cite{Zayed2017}.

\begin{figure}[b]
\includegraphics[width=84mm]{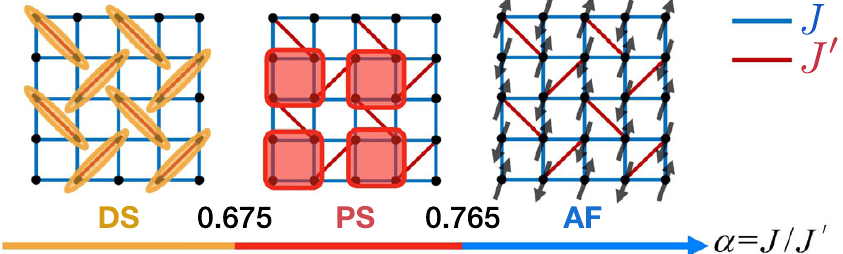}
\caption{Schematic $T=0$ phase diagram of the SS model \cite{koga00,Corboz2013}.} 
\label{fig:ss}
\end{figure}

The unpaired $S=1/2$ Cu spins of SrCu$_2$(BO$_3$)$_2$ form layers of orthogonal dimers
\cite{Kageyama1999,Miyahara1999}. The two dominant Heisenberg exchange couplings $J_{ij}{\bf S}_i \cdot {\bf S}_j$ realize the Shastry-Sutherland (SS)
model \cite{shastry81}, illustrated in Fig.~\ref{fig:ss}, with intra- and inter-dimer values $J' \approx 75$ K and $J \approx 45$ K, respectively. 
The SS model has an exact dimer-singlet (DS) ground state for $0 \le \alpha=J/J' \alt 0.68$ \cite{shastry81,koga00,Corboz2013} and for  
$\alpha \to \infty$ it reduces to the Heisenberg AF \cite{Manousakis1991}. There is a PS phase between the DS and AF phases, 
at $\alpha \in [0.68,0.76]$ \cite{koga00,Corboz2013}. 

At ambient pressure the properties of SrCu$_2$(BO$_3$)$_2$ agree well with the SS model in the DS phase \cite{Kageyama1999,Miyahara1999}. AF order 
has been observed at $P \approx 4$ GPa \cite{Zayed2017}, close to a tetragonal--monoclinic structural transition \cite{Loa2005,Zayed2014,Haravifard2014}. 
Since the Mermin-Wagner theorem prohibits $T>0$ magnetic order in a 2D spin-isotropic system, the AF order should be due to weak inter-layer couplings
(and possibly some spin anisotropy). A 2D SS description of the quantum phase transitions is still relevant, and the simplest explanation of the behavior
is that $\alpha$ increases with $P$ \cite{koga00,Haravifard2012,Zayed2017}. Then it should also be possible to stabilize the PS phase of the SS model
at intermediate $P$ and low $T$. Breaking a discrete two-fold Ising ($Z_2$) symmetry, corresponding to two equivalent plaquette patterns,
PS order can appear at $T>0$ already in an isolated layer. 

Following indications from NMR of an intermediate phase with broken spatial symmetry \cite{Waki2007,Haravifard2016}, inelastic neutron scattering
revealed an excitation attributed to a PS state \cite{Zayed2017}. The mode was only detected at $P=2.15$ GPa, and recently an alternative scenario
with no PS phase was proposed \cite{Boos2019}. Here we argue that the PS phase exists adjacent to a previously not observed AF phase below $4$ K
and $P = 3$ - $4$ GPa.

\begin{figure}
\includegraphics[width=66mm]{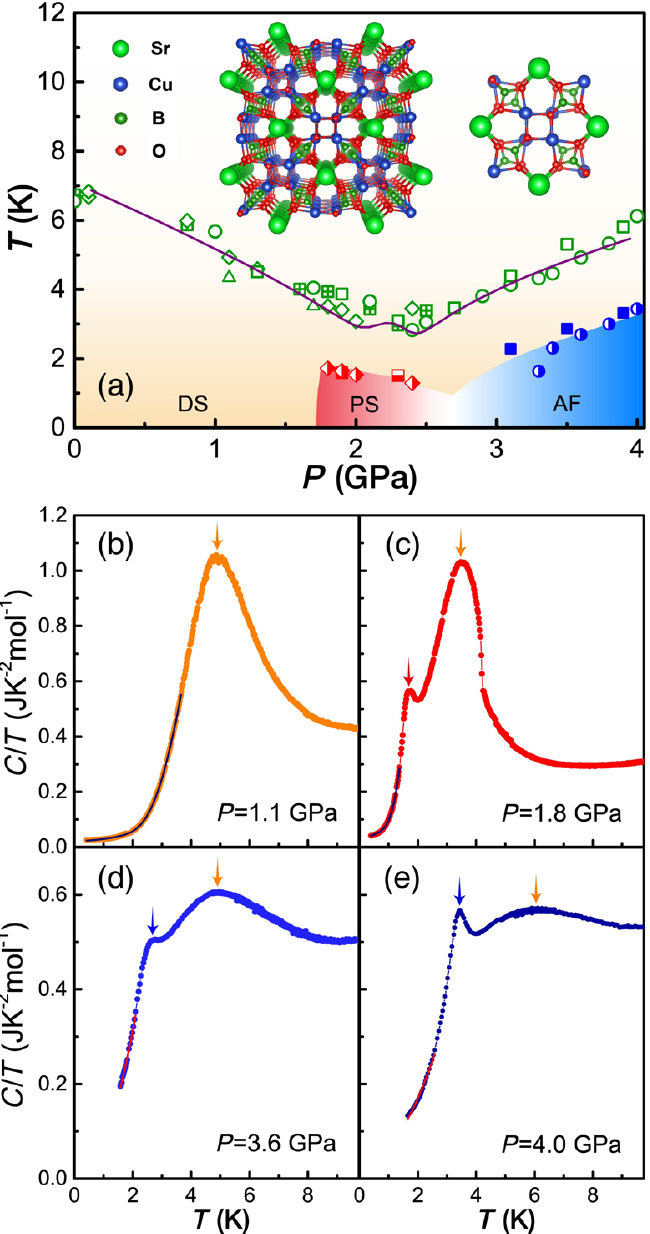}
\caption{(a) Phase diagram of SrCu$_2$(BO$_3$)$_2$ (crystal structure in the inset) from high pressure $C(T)$ measurements. 
Examples of $C(T)/T$ curves are given in (b)-(e), where the orange arrows indicate the hump location $T_h$. The green symbols in (a) 
mark $T_h$ in several samples and the purple curve shows results for the 
$20$-spin SS model with couplings close to those of Ref.~\cite{Zayed2017}; $J'(P)=[75-8.3P/{\rm GPa}]$ K and $J(P)=[46.7-3.7P/{\rm GPa}]$ K. 
For $P \approx 1.7$ - $2.4$ GPa a second peak at lower $T$, marked with a red arrow in (c); it indicates the transition into the PS phase. Upon 
further compression, the system first enters a regime where the experiments cannot reach sufficiently low $T$ to observe the second peak. 
The peak is again detectable around $3$ GPa and is marked with blue arrows in (d),(e). It becomes more prominent with increasing $P$, suggesting
\cite{Sengupta2003} AF order due to weak inter-layer couplings. The phase boundaries extracted from the second peak are indicated by half-filled 
red squares and diamonds (PS phase) and blue filled squares and half-filled circles (AF phase). The low-$T$ data in (b,c) are fitted (black curves) 
to the form $C/T=a_0 + a_1T^2 +(a_2/T^3) {\rm e}^{-\Delta/T}$ \cite{Kageyama2000}, giving gaps $\Delta$
shown in Fig.~\ref{fig:gap}(a). In (d,e) fits are shown (red curves) without gap term; $C/T=a_0 + a_1T^2$.}
\label{fig:phases}
\end{figure}

{\it Experiments}.---We
have performed high-pressure heat capacity ($C$) measurements on SrCu$_2$(BO$_3$)$_2$ single crystals. With support of simulations of quantum spin
models, we have for the first time extracted a $(P,T)$ phase diagram, Fig.~\ref{fig:phases}(a), in the range of $P$ and $T$ where the SS model should
be relevant. Six different samples were studied, and $C(T)$ was measured from room temperature down to $1.5$ K or $0.4$ K at several pressures (using
two different types of cryostats and pressure cells; see Supplemental Material, SM \cite{sm}). Consistent results were obtained among all these measurements. 
In Fig.~\ref{fig:phases}(b-e) we show typical results for $C(T)/T$ in the different pressure regions. In SM \cite{sm} we discuss data for $P > 4$ GPa,
where the SS description is no longer valid.

We identify two main low-$T$ features in $C(T)/T$: There is always a broad maximum that we will refer to as the hump. Starting
at $P \approx 1.7$ GPa, a smaller peak emerges at lower $T$ and prevails up to $2.4$ GPa. We will argue that this peak signals the PS
transition. Upon further increasing $P$, the small peak is no longer detected at temperatures accessible in the experiments. A broader hump
appears between $3$ and $4$ GPa, below which there is a peak at $T \approx 2$ - $3.5$ K that we interpret as an AF transition. AF order was previously
detected only at $P>4$ GPa up to $T \approx 120$ K \cite{Zayed2017}. This high-$T$ phase is different from the new low-$T$ AF phase---see SM \cite{sm},
where we also discuss a new transition at $T \approx 8$ K for $P > 4$ GPa.

The $C/T$ hump is known from studies at ambient pressure \cite{Kageyama2000}, where it arises from the correlations leading to the dimer singlets as 
$T \to 0$. As shown in Fig.~\ref{fig:phases}(a), the hump temperature $T_h(P)$, including the minimum at $P \approx 2.5$ GPa, agrees remarkably well 
with exact diagonalization (ED) results for the SS Hamiltonian on a $20$-site lattice (see SM \cite{sm}) with $P$ converted to $\alpha$ by linear 
forms $J(P)$, $J'(P)$ \cite{Zayed2017}. The hump width also agrees well with the SS model [see Fig.~S5].

In the 2D Heisenberg model the hump appears at $T\approx J/2$ \cite{Sengupta2003} where
strong AF correlations build up. In general, the hump indicates a temperature scale where correlations set in that remove significant
entropy from the system.  The $T_h(P)$ minimum can be regarded as the point of highest frustration, with the energy scale being lowered
due to the two competing couplings (see also Refs.~\cite{Prelovsek2018,Li2018}). The peak that we associate with PS ordering appears
in this pressure region, suggesting singlet formation driven by strong frustration.

If the putative AF ordering below $T = 4$ K for $P \approx 3$ - $4$ GPa is the result of weak inter-layer couplings $J_\perp$, the observed hump-peak
separation is expected, as the hump present for an isolated layer is not affected much by a small $J_\perp$ and $T_{\rm AF} \to 0$ as $J_\perp \to 0$.
Moreover, the ordering peak vanishes as $J_\perp \to 0$, because most of the entropy has been consumed by 2D correlations before 3D long-range order
sets in. Our results at $3.6$ GPa and $4.0$ GPa compare favorably with quantum Monte Carlo (QMC) calculations of weakly coupled Heisenberg
layers \cite{Sengupta2003} with $J_\perp/J_{\rm 2D} \approx 0.01$ - $0.02$. In the SS system $J_{\rm 2D}$ is an effective 2D AF coupling smaller 
than both $J$ and $J'$ (because of frustration). The more prominent low-$T$ peak and higher $T_{\rm AF}$ at higher $P$ should be a consequence of $\alpha$
increasing, likely in combination with an increase of $J_\perp$. The low-$T$ peak becomes harder to discern as $P$ is decreased down to $3$ GPa, where 
$T_c$ is lower \cite{Sengupta2003}. Unfortunately, above $2.4$ GPa we are restricted to $T\ge1.5$ K and cannot track the PS and AF transitions within 
the white region in Fig.~\ref{fig:phases}(a).

\begin{figure}[t]
\includegraphics[width=80mm]{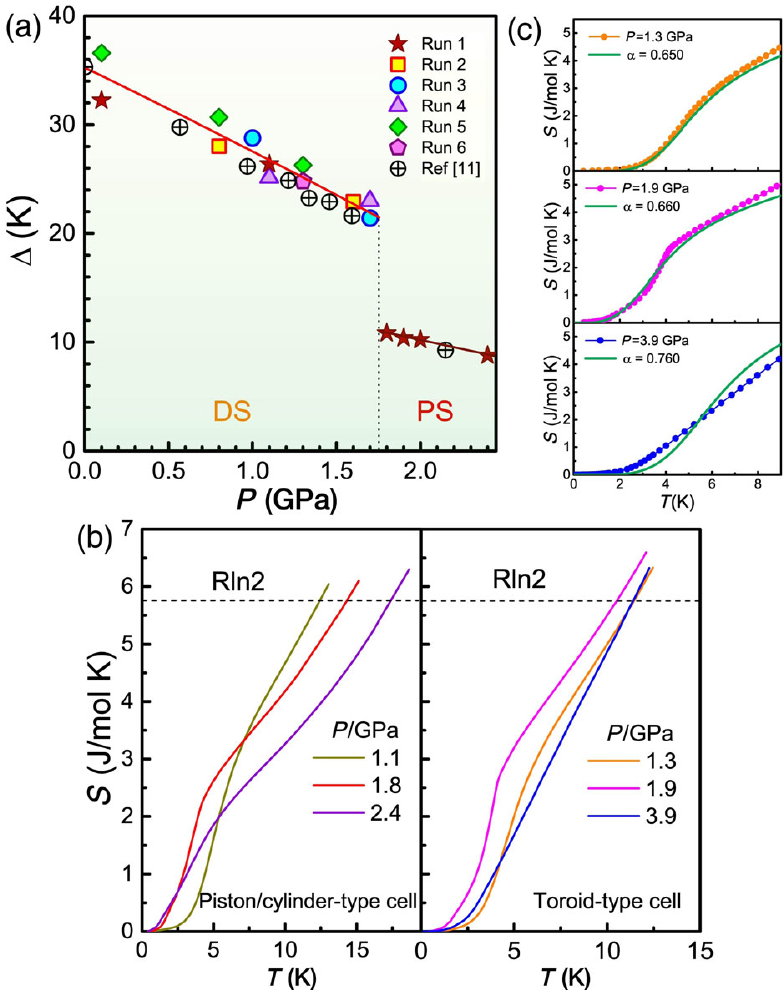}
\caption{(a) Pressure dependent gaps extracted from low-$T$ fits to $C(T)/T$ [Fig.~\ref{fig:phases}(b) and \ref{fig:phases}(c)] for six different experimental 
runs, compared with the neutron scattering results \cite{Zayed2017}; the $\oplus$ mark at $P = 2.15$ GPa refers to the low-energy excitation in Fig.~1 of 
Ref.~\cite{Zayed2017}. (b) Examples of the entropy obtained by integrating $C/T$ over $T$ [using fits such as those in Fig.~\ref{fig:phases}(b-e) at low $T$]. 
(c) Results from the Toroid-type cell compared with the 20-spin SS model with couplings given by the formula used in Fig.~\ref{fig:phases}(a). The results are 
normalized to the unit cell containing a Cu dimer; thus $S(T \to \infty) = 2R\ln(2)$ [twice the dashed line value in (b)].}
\label{fig:gap}
\end{figure}

Our identification of the phases partially rely on the low-$T$ tails in $C/T$.
Up to $P = 2.4$ GPa we extracted the gap by fitting $C(T)/T$ to an exponential form plus terms accounting for the
heater, wires, and phonons [Fig.~\ref{fig:phases}(b,c)]. The $P$ dependent gaps [Fig.~\ref{fig:gap}(a)] are in excellent agreement with previous
works using different methods. The gap is suddenly reduced by a factor of two at $1.7$ GPa, showing that the DS--PS transition
is first-order, as in the SS model \cite{koga00,Corboz2013}. In our proposed AF phase $C(T)/T$ can be fitted [Fig.~\ref{fig:phases}(d,e)]
without a gap.

Fig.~\ref{fig:gap}(b) shows examples of the entropy obtained by integrating $C(T)/T$ in the DS, PS, and AF states.
Data sets from experiments with the two different pressure cells exhibit consistent trends. Comparing the results with the SS model [Fig.~\ref{fig:gap}(c)] 
confirms  that the features in $C/T$ below $T \approx 8$ K predominantly originate from the Cu $S=1/2$ spin network. The agreement between the experimental 
and theoretical results is striking at $P=1.3$ and $1.9$ GPa, where the system is gapped. At $P=3.9$ GPa the SS model still captures the overall magnitude 
of the entropy, though the AF state can naturally not be fully reproduced by a small 2D cluster.

{\it Modeling}.---Ideally, we would like to compare the experiments with the SS model supplemented by weak 3D couplings. However, calculations
at low $T>0$ in the PS and AF phases require much larger lattices than those accessible to ED, and other numerical techniques are also very
challenging \cite{Prelovsek2018,Li2018}. To investigate generic aspects of the PS and AF transitions, we instead study a 'J-Q' model amenable to 
large-scale QMC simulations. The model was proposed \cite{Sandvik2007} for studies of deconfined quantum criticality \cite{senthil04,shao16}, 
and recently a 'checker-board' variant (CBJQ model) was deviced for realizing the PS--AF transition \cite{Zhao18}.

\begin{figure*}[t]
\includegraphics[width=170mm]{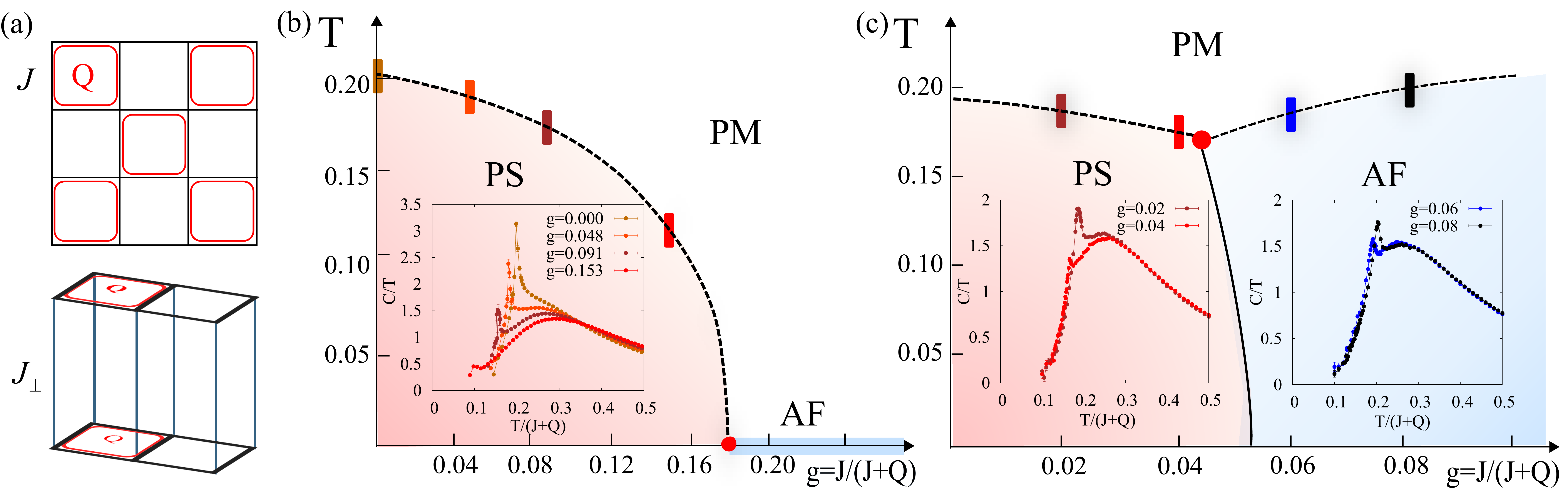}
\caption{(a) In the CBJQ model the SS $J'$ exchange [Fig.~\ref{fig:ss}] is replaced by 
four-spin interactions $-Q({\bf S}_i \cdot {\bf S}_j-1/4)({\bf S}_k \cdot {\bf S}_l-1/4)$, where $ij$ and $kl$ form edges 
(both horizontally and vertically) of the plaquettes 
with red squares \cite{Zhao18} 
. Intra- and inter-layer Heisenberg couplings are denoted by $J$ and $J_\perp$, respectively.
(b) $J_\perp=0$ phase diagram with $g=J/(J+Q)$. The PS--AF quantum-critical point is at $g_c \approx 0.179$ and there is
AF order only at $T=0$. The inset shows $C/T$ for lattices up to size $256^2$ at $g<g_c$. The hump-peak separation increases and the area 
under the peak decreases as $g \to g_c$. (c) Phase diagram at $J_\perp/(J+Q)=0.1$ obtained with up to $48\times 48\times 24$ spins. 
The insets show examples of $C(T)/T$ curves.}
\label{fig:models}
\end{figure*}

The $Q$ interactions of the CBJQ model [Fig.~\ref{fig:models}(a)] compete against AF order and lead to an unusual transition versus
$g=J/Q$ where the scalar (Z$_2$) PS and O($3$) AF order parameters combine into an O($4$) vector  \cite{Zhao18}. Even though the CBJQ and SS
models are different at the lattice level, one can expect universal large-scale physics. Thus, SrCu$_2$(BO$_3$)$_2$ may also realize emergent O(4)
symmetry---if indeed it hosts a low-$T$ PS--AF transition dominated by 2D quantum fluctuations. Here we do not address the issue of emergent
symmetry directly, but focus on the thermodynamics. The models and QMC technique are further discussed in SM \cite{sm}.

Fig.~\ref{fig:models}(b) shows $C/T$ for different coupling ratios $g$ in the 2D CBJQ model. The peak signaling the PS transition  gradually separates
from a hump as $g$ increases, at the same time shrinking as there is less entropy associated with the phase transition. The short-range correlations
signaled by the hump are predominantly AF in nature but also reflect the formation of singlets on the plquettes before the collective ordering of those
singlets. The clear hump-peak separation and the small ordering peak when $g \approx g_c$ are signatures of strong 2D quantum fluctuations of the PS order
and are strikingly similar to our observations in SrCu$_2$(BO$_3$)$_2$ [Fig.~\ref{fig:phases}(c)].

To study AF order at $T>0$ we introduce inter-layer couplings $J_\perp$ [Fig.~\ref{fig:models}(a)]. Fig.~\ref{fig:models}(c) shows the phase diagram for
a moderately small $J_\perp$ along with scans of $C/T$. We observe a hump-peak structure close to the PS--AF transition; in particular the behavior in
the vicinity of the AF transition is similar to the results for SrCu$_2$(BO$_3$)$_2$, thus supporting our conclusion of an AF phase in the material
at $P = 3$ - $4$ GPa.

Our SS model fit to the experimental hump in Fig.~\ref{fig:phases}(a) gives $\alpha \approx 0.665$ at the DS--PS transition,
close to the transition point in the SS model. In the white region in Fig.~\ref{fig:phases}(a) we have $\alpha \approx 0.69-0.71$, which is
smaller than $\alpha \approx 0.76$ at the PS--AF transition in the SS model. Inter-layer exchange interactions will enhance the AF correlations 
and should shift the boundary of the AF phase in the way observed. 
An analogous effect of $J_\perp$ on the PS--AF transition in the CBJQ model is seen in Fig.~\ref{fig:models} for $J_\perp=0.1$, and even for 
$J_\perp=0.01$ we still see a shift of $g_c$ by $\approx 10\%$, as shown in SM \cite{sm}. We are not aware of any estimates of $J_\perp$ 
in SrCu$_2$(BO$_3$)$_2$, but our results show that the quantitative effects of this coupling on the phase diagram should not be neglected, even though 
the low-$T$ quantum fluctuations remain predominantly 2D in nature.

{\it Discussion}.---The singlets in the PS phase of SrCu$_2$(BO$_3$)$_2$ may form on the dimer plaquettes \cite{Zayed2017}, not on the empty plaquettes as
in the SS model \cite{Corboz2013}. It was recently proposed that the state is not even a 2-fold degenerate PS state with a symmetry-breaking transition,
but a state resulting from an orthorombic distortion \cite{Boos2019}. This would be consistent with NMR results showing two kinds of dimers below $3.6$ K at $2.4$
GPa \cite{Waki2007}. In our experiments, the hump in $C(T)/T$ for $P$ between $1.7$ and $2.4$ GPa is close to this NMR splitting temperature, and the hump 
also some times has a small jump on its right side, as in Fig.~\ref{fig:phases}(c). Our modeling shows clearly that the hump is a consequence of short-range 
correlations and does not originate from a phase transition, but the jump could still be due to a weak orthorombic transition (which might even be driven by 
the spin correlations). Given overall small effects on $C(T)$, such a transition (if it exists) may not change the couplings as much as suggested by Boos 
et al.~\cite{Boos2019}, who also agree that the PS state can still exist with a very weak orthorombic distortion \cite{Boos2019}. Their alternative 
quasi-1D state would not undergo any further phase transition at lower $T$, contradicting the clear peaks we find for intermediate pressures at $T \approx 2$ K. 
The quasi-1D scenario was in part motivated by the gap decreasing with $P$ [as we also have found; Fig.~\ref{fig:gap}(a)] \cite{Boos2019} (see also
Ref.~\cite{Nakano2018} for SS model ED results). However, the gap calculations are subject to approximations, and even small interactions beyond the SS 
model (e.g., 3D couplings) may play a role as well in the gap evolution in SrCu$_2$(BO$_3$)$_2$. Recent ESR experiments at $P \approx 2$ GPa were explained 
with a PS phase surviving in the presence of a pressure-induced weak distortion \cite{Sakurai2018}. 

In an alternative scenario, the $C/T$ peak at $T \approx 2$ K could reflect an orthorombic transition, with the NMR splitting brought to higher $T$ 
by magnetic-field effects (if the orthorombic transition is sensitive to spin correlations). However, it has also been argued from other experiments 
that there is no structural transition at $P \approx 2$ GPa \cite{Haravifard2012,Sakurai2018}. It would be useful to repeat the NMR experiments for a wider 
range of pressures and study field effects systematically. It is also not completely clear whether the singlets in SrCu$_2$(BO$_3$)$_2$ really form on the 
dimer plaquettes, as calculations of the spectral signatures have only been calculated on very small systems \cite{Zayed2017} or in perturbative schemes 
\cite{Boos2019} that may not sufficiently account for the complexities of the PS quantum fluctuations. 

The simplest scenario is that the phase boundaries of the low-$T$ PS and AF phases of SrCu$_2$(BO$_3$)$_2$ can be explained by the 2D SS model with weak 3D interlayer
couplings. The existence of the new low-$T$ AF state argued here 
resolves a puzzling aspect of the phase diagram \cite{Zayed2017} that had not been emphasized previously: 
a high-$T$ AF transition, with $T_{\rm HT} \approx 120$ K, is inconsistent with SS couplings $J,J' \ll T_{\rm HT}$ and the frustration that further reduces the 
effective magnetic energy scale $J_{\rm 2D}$. The deconfined quantum-criticality scenario for the PS--AF transition would be unlikely under these circumstances. In
contrast, $T_{\rm AF} < 4$ K found here is compatible with the SS model and $J_\perp \ll J,J'$. Although we were not able to track the phase boundaries in the region 
$P \approx 2.4$ - $3.1$ GPa [Fig.~\ref{fig:phases}(a)], the most natural scenario is a direct PS--AF transition below $T \approx 1$ K. This transition should be 
weakly first-order, related to the deconfined quantum-criticality scenario \cite{senthil04,shao16,Lee2019} and with an emergent O(4) symmetry of the two order 
parameters \cite{Zhao18,Serna2018} if the 3D couplings are sufficiently weak. Our study has established the $(P,T)$ region in which to further 
investigate this physics experimentally.

It will be important to confirm the magnetic structure of the new low-$T$ AF phase by neutron scattering---the previous experiments in this pressure range
did not reach down to the transition temperatures we found here \cite{Zayed2017}. A Raman spectroscopy study reported after the completion of our 
work \cite{Bettler2018} has already detected correlations compatible with AF ordering at pressures similar to Fig.~\ref{fig:phases}(a). It would also be 
interesting to investigate magnetic field effects. Further model calculations should test the stability of the emergent O(4) symmetry \cite{Zhao18,Serna2018} 
and other aspects of the PS--AF transition related to deconfined quantum criticality beyond the strict 2D limit. 

\begin{acknowledgments}
{\it Acknowledgments.}---The 
research at Chinese institutions was supported by the National Key Research and Development Program of China (Grants No.~2017YFA0302900, 2016YFA0300300,
2016YFA0300502, 2017YFA0303103), the NSF of China (Grants No.~11427805, U1532267, 11604376, 11874401, 11674406, 11874080, 11421092, 11574359, 11674370),
and the Strategic Priority Research Program (B) of the Chinese Academy of Sciences (Grants No.~XDB25000000, XDB07020000, XDB28000000). The work in Boston
was supported by the NSF under Grant No.~DMR-1710170 and by a Simons Investigator Award. J.G. also acknowledges funding from the Youth Innovation Promotion 
Association of the Chinese Academy of Sciences (Grant No.~2019008) and V.A.S. acknowledges the support of RFBR grant No.~18-02-00183. We thank the Center 
for Quantum Simulation Sciences in the Institute of Physics, Chinese Academy of Sciences and the Tianhe-1A platform at the National Supercomputer Center 
in Tianjin for their technical support and generous allocation of CPU time. Some of the numerical calculations were carried out on the Shared Computing
Cluster managed by Boston University's Research Computing Services.
\vskip5mm

\end{acknowledgments}

\begin{widetext}

\newpage
  
\begin{center}  

\section{Supplemental Material}

{\bf\large \noindent Quantum phases of SrCu$_2$(BO$_3$)$_2$ from high-pressure thermodynamics}
\vskip5mm

{\noindent
Jing Guo,$^1$ Guangyu Sun,$^{1,2}$ Bowen Zhao,$^{3}$ Ling Wang,$^{4,5}$ Wenshan Hong,$^{1,2}$ Vladimir A. Sidorov,$^{6}$ Nvsen Ma,$^{1}$ Qi Wu,$^{1,2}$ \\
Shiliang Li,$^{1,2,7}$ Zi Yang Meng,$^{1,7,8,*}$ Anders W. Sandvik,$^{3,1,*}$ and Liling Sun $^{1,2,7,*}$}
\vskip3mm

{\it
{$^1$Beijing National Laboratory for Condensed Matter Physics \\ and Institute of Physics, Chinese Academy of Sciences, Beijing 100190, China} \\
{$^2$School of Physical Sciences, University of Chinese Academy of Sciences, Beijing 100190, China} \\
{$^3$Department of Physics, Boston University, 590 Commonwealth Avenue, Boston, Massachusetts 02215, USA} \\
{$^4$Beijing Computational Science Research Center, 10 East Xibeiwang Road, Beijing 100193, China} \\
{$^5$Zhejiang Institute of Modern Physics, Zhejiang University, Hangzhou 310027, China} \\
{$^6$Vereshchagin Institute for High Pressure Physics, Russian Academy of Sciences, 108840 Troitsk, Moscow, Russia} \\
{$^7$Songshan Lake Materials Laboratory, Dongguan, Guangdong 523808, China} \\
{$^8$Department of Physics and HKU-UCAS Joint Institute of Theoretical and Computational Physics, \\ The University of Hong Kong, Pokfulam Road, Hong Kong, China}\\}
\vskip1mm
$^*$ e-mail: zymeng@iphy.ac.cn, sandvik@bu.edu, llsun@iphy.ac.cn

\end{center}
\vskip5mm

Here we present additional supporting results for the findings in the main paper. In Sec.~1 we detail the crystal growth procedures.
In Sec.~2 we show complete data sets for all the measurements
taken for three different SrCu$_2$(BO$_3$)$_2$ samples with the two different high-pressure cells. In Sec.~3 we discuss the phase diagram of
SrCu$_2$(BO$_3$)$_2$ at higher pressures than considered in the main text, extending the results to the region $P = 4.2$ to $4.9$ GPa
where two phases appear that are not related to the SS description. In Sec.~4 we discuss finite-size effects in the ED
results for the SS model and present $C(T)/T$ curves for several values of the coupling ratio $\alpha$. In Sec.~5 we provide details of the QMC
results and finite-size scaling analysis for the CBJQ model in two and three dimensions and provide additional results supporting our
conclusions regarding the role of the inter-layer coupling $J_\perp$.\null\vskip5mm

\end{widetext}

\vskip5mm

\setcounter{page}{1}
\setcounter{equation}{0}
\setcounter{figure}{0}
\renewcommand{\theequation}{S\arabic{equation}}
\renewcommand{\thefigure}{S\arabic{figure}}

\subsection{1. Single crystal growth}

High-quality single crystals of SrCu$_2$(BO$_3$)$_2$ were grown by a traveling floating-zone method similar to what has been
reported in the literature previously~\cite{Dabkowska2007}. The mixture of SrCO$_3$, CuO and B$_2$O$_3$ in stoichiometric proportions
was ground and heated at 780 $^\circ$C for 24 hours. After repeating these procedures at 800 and 820 $^\circ$C, the powders were pressed 
hydrostatically into a cylindrical rod with diameter of about 7 mm. The rods were annealed in flowing oxygen at 1000 $^\circ$C 
for 12 hours. The crystals were thereafter grown in 4 atm of oxygen at a speed of 0.5 mm/h, until the single-crystal rods reached
a length of approximately 50mm. From these rods, small pieces of size on the order of $1\times 1 \times 0.2$ mm were chipped off and
polished for smoothness.

\subsection{2. High pressure heat capacity measurements \\ in two types of pressure cells}

In this study, two types of high pressure cells were employed for the heat capacity measurements due to the restriction of the inner space of
our extremely-low temperature system. A piston/cylinder-type high pressure cell with Daphne 7373 oil as pressure transmitting medium was used for
the measurements up to $2.4$ GPa for temperatures down to $0.4$ K. The larger Toroid-type high pressure cell \cite{Petrova2005} with glycerin/water
(3:2) liquid as the pressure transmitting medium was adopted for the measurements up to $\approx 5$ GPa at temperatures down to $1.5$ K.
The pressure was determined by the pressure dependent superconducting $T_c$ of a piece of Pb that was placed in the Teflon capsule together
with the sample \cite{Eiling1981}.

Single-crystal SrCu$_2$(BO$_3$)$_2$ samples with dimensions of about $0.9 \times 0.9 \times 0.18$ mm$^3$ and $0.8 \times 0.4 \times 0.15$ mm$^3$ were
used for the piston/cylinder-type and the Toroid-type high pressure cell, respectively. Platinum wires of diameter $25$ $\mu$m were spot-welded to the
ends of the heater and its resistance was a few Ohms. Constantan was used for the heater. This is a convenient heater  material because its resistivity
has only has a weak
temperature dependence. The room temperature resistance $R$ of the heater was determined by measuring its length under microscope and using the known
resistance per unit length of our wire measured separately. An (Au$_{0.07}$Fe)-chromel thermocouple was glued to the opposite side of the crystal. A
sine wave AC excitation current $I$ at frequency $f$ was applied to the heater and the resulting temperature oscillations $\Delta T$ of the sample
temperature at frequency 2$f$ was detected by the thermocouple amplified by an SR554 preamplifier and measured by an SR830 lock-in amplifier. As
the input power $P$ is known ($P = I^2R$) we can calculate the product ($P/f\Delta T$) which is proportional to the heat capacity at the optimal
measuring frequency \cite{Eichler1979,Kraftmakher2002}. The optimal frequency of the AC-power input was varied on cooling to maintain quasi-adiabatic
conditions needed for correct calorimetry measurements.

\begin{figure}[t]
\includegraphics[width=78mm, clip]{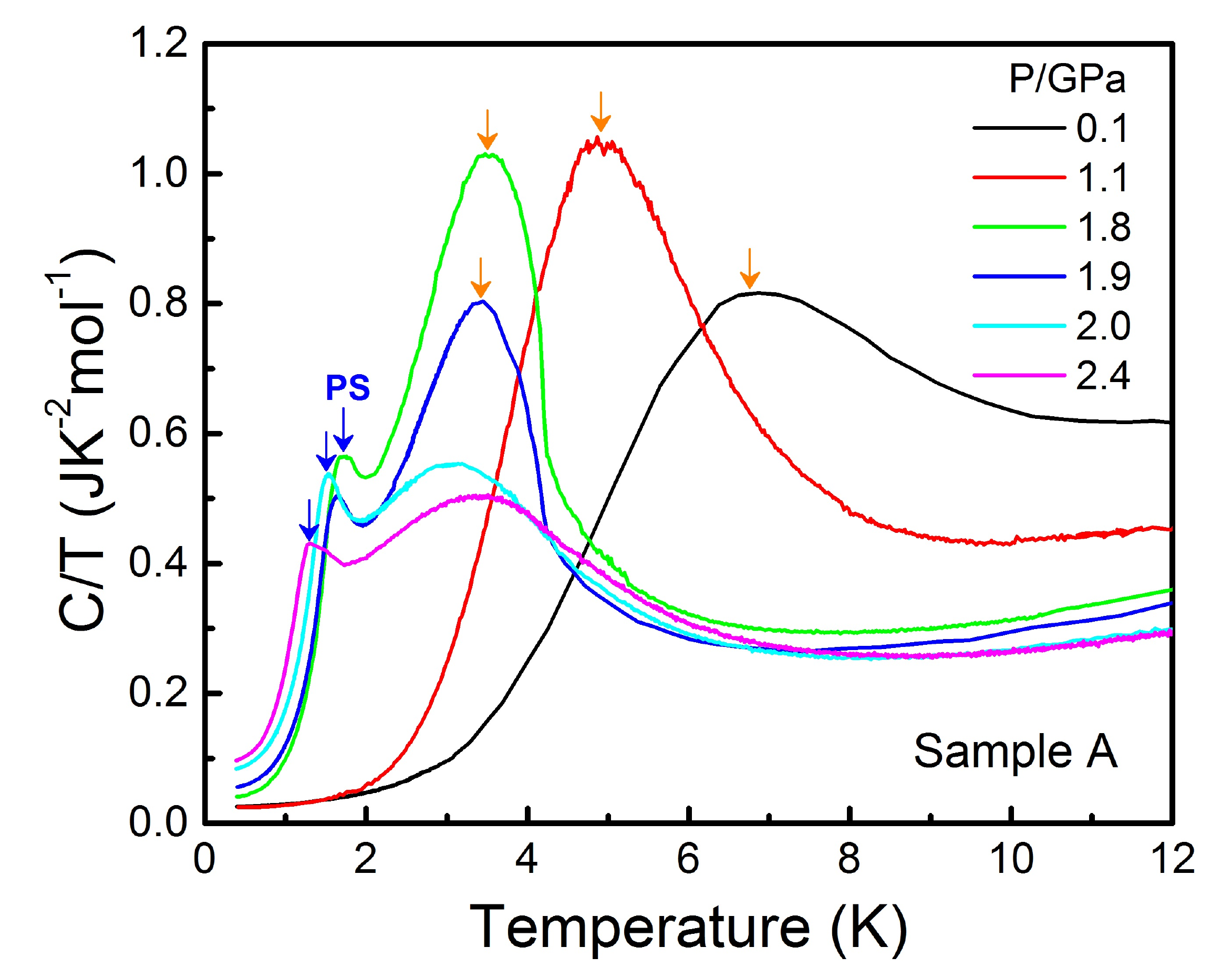}
\caption{Temperature dependence of $C/T$ for SrCu$_2$(BO$_3$)$_2$ sample A measured at different pressures with the piston/cylinder-type high pressure cell
for temperatures down to $T=0.4$ K. The arrows indicate the hump temperature $T_h$ and the peak associated with ordering into the PS state.}
\label{fig:s1}
\end{figure}

\begin{figure}[t]
\begin{center}
\includegraphics[width=78mm, clip]{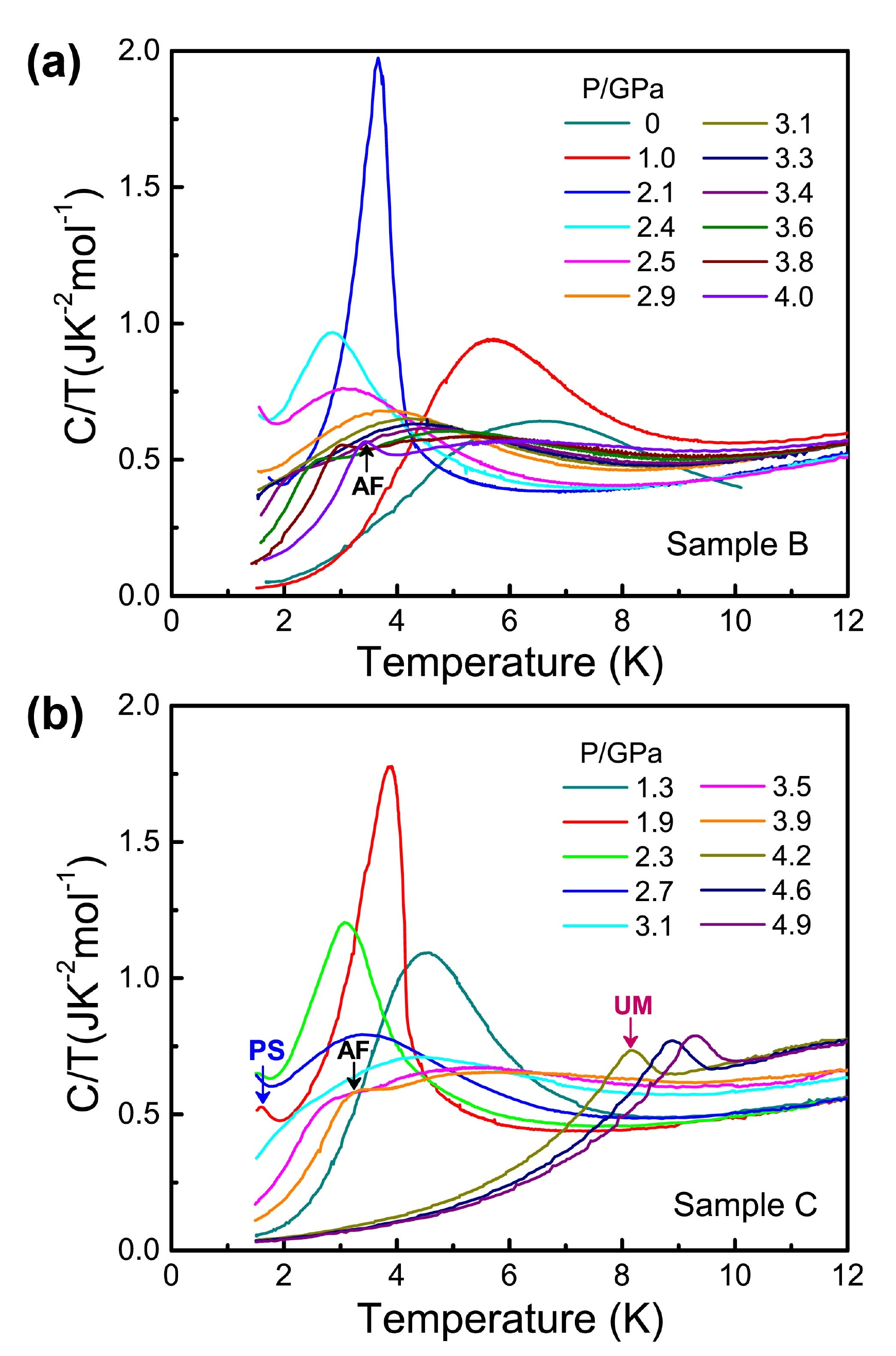}
\caption{Heat capacity $C/T$ as a function of temperature for sample B (a) and sample C (b) measured using the Toroid-type high-pressure
cell for temperatures down to $T=1.5$ K. In addition to the PS and AF phase transitions discussed in the main text, we find a transition
(peaks marked UM) into an unknown magnetic state.}
\label{fig:s2}
\end{center}
\vskip-3mm
\end{figure}

Difficulties for quantitative high-pressure AC-calorimetry arise from the presence of the pressure transmitting medium surrounding
the sample, which acts as an effective addenda together with the heater, glue, and part of the connecting wires adjacent to the sample. The
contribution of the pressure transmitting glycerin-water medium was estimated from the measured value of the reduction of the specific heat
(in $\mu$J/K) at the glass transition in this liquid upon cooling. Separate experiments with liquid alone in the pressure cell give a map of
$C(P,T)$ for the glycerin-water mixture and allow us to estimate its contribution to the total heat capacity measured by the Toroid-type
pressure cell. For the Daphne 7373 liquid this information is not available. The contribution of Daphne 7373 oil surrounding the sample in the
piston/cylinder pressure cell was instead estimated from AC-calorimetry measurements of the sample-heater-thermocouple assembly at ambient pressure
down to $0.4$ K in vacuum and the same assembly in Teflon capsule filled with Daphne 7373 liquid. The results of these experiments allow us to 
calibrate our measurements to the previously published ambient-pressure $C(T)$ curve for of SrCu$_2$(BO$_3$)$_2$ \cite{Kageyama2000b}. We assume
that this calibration is satisfactory up to 2.4 GPa.

Although a major part of the addenda related with heater, connecting wires and glue is removed by this procedure, there are still
some remaining contributions to $C(T)$. That is why in the fits of the low-temperature specific heat in Figs.~\ref{fig:phases}(b,c) the $T$ linear and cubic
terms are present in addition to the exponential term originating from from the dominant magnetic specific heat of the SrCu$_2$(BO$_3$)$_2$ sample.
The gaps obtained from these fits do not depend significantly on the presence of residual addenda contributions.

\begin{figure*}[t]
\includegraphics[width=140mm]{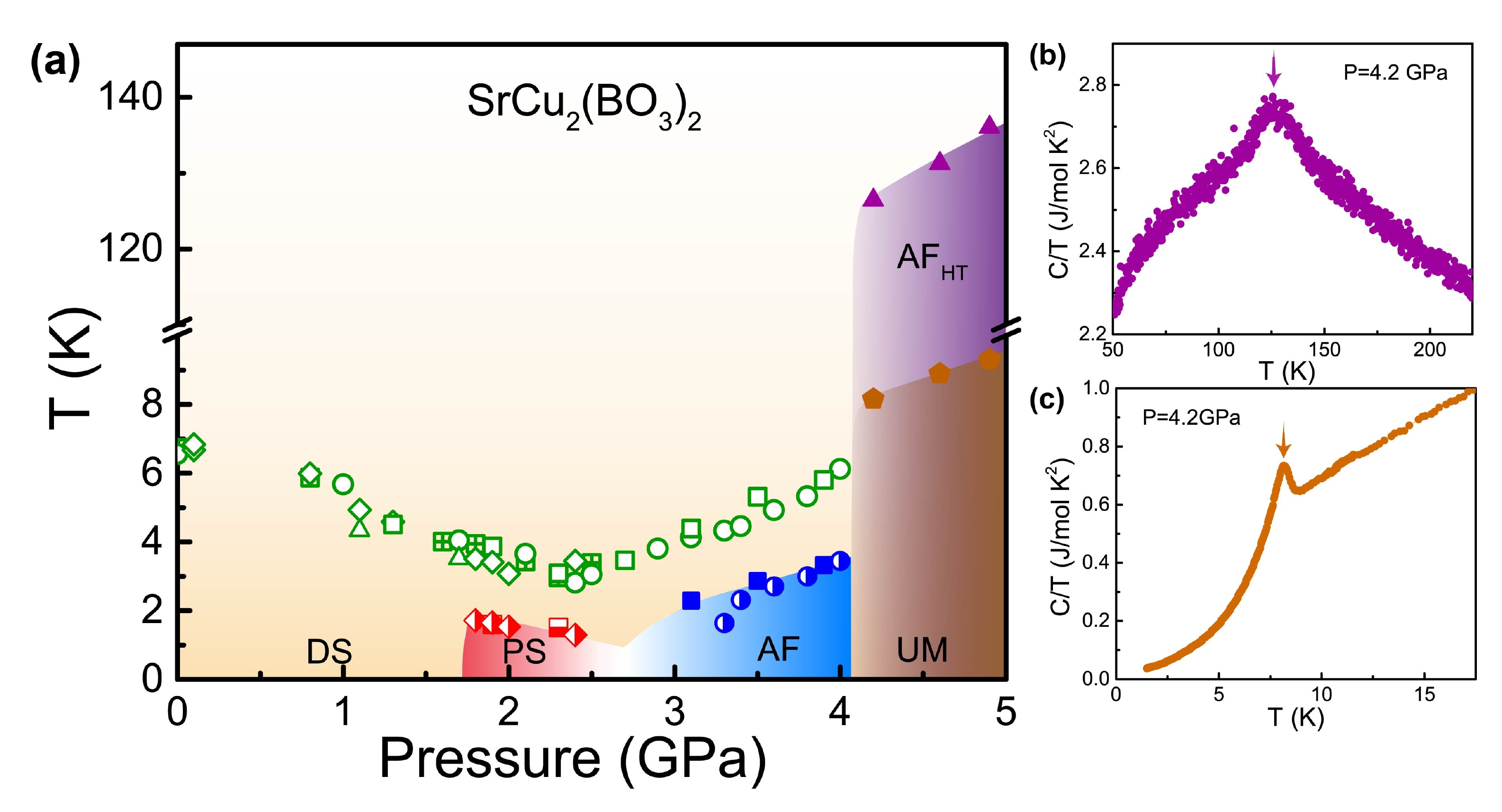}
\caption{Extended ($P,T$) phase diagram of SrCu$_2$(BO$_3$)$_2$. In (a) the green open symbols stand for the characteristic $C/T$ hump temperature 
obtained from independent runs, the half-filled diamonds and squares represent the transition temperature of the PS phase, the blue solid squares
and half-filled circles stand for the onset temperature of the AF state. The brown pentagons represent the onset temperature of the new UM state.
The purple triangles stand for the onset temperature of the previously known \cite{Zayed2017} high-temperature AF$_{\rm HT}$ state. Examples of
the peaks in $C/T$ associated with the AF$_{\rm HT}$ and UM transitions are shown in (b) and (c), respectively.}
\label{fig:s3}
\end{figure*}

In the main text we presented typical $C(T)/T$ curves in Figs.~\ref{fig:phases}(b-e). Here we show a larger set of curves obtained with
the two different pressure cells (described in Methods).

Figure \ref{fig:s1} shows the high pressure heat capacity measurements obtained by using the piston/cylinder-type high pressure cell for pressures
from $0.1$ GPa to up to $2.4$ GPa and temperatures ranging from $0.4$ K to $12$ K. It can been seen that, at $0.1$ GPa, the plot of $C/T$ versus temperature
displays a hump behavior which has been considered to be related to the formation of dimer single state, in good agreement with its ambient-pressure behavior
reported previously \cite{Kageyama2000}. The hump is found to shift to lower temperature initially with increasing pressure below $2.0$ GPa and then moves
to higher temperature with further compression. Remarkably, upon increasing $P$ to $1.8$ GPa (green curve in Fig.~\ref{fig:s1}), a smaller peak appears
at $1.7$ K and it systematically shifts to lower temperature when $P$ is increased to $2.4$ GPa. The hump and peak temperatures are marked by the open green
and half-filled red diamonds, respectively, in Fig.~\ref{fig:phases}(a) in the main text.

To reveal the behavior of SrCu$_2$(BO$_3$)$_2$ at higher pressure, we carried out heat capacity measurements in a Toroid-type pressure cell which allows
us to apply pressure up to $\approx 5$ GPa. Figures \ref{fig:s2}(a) and \ref{fig:s2}(b) display the results from two independent runs with two different
single-crystal samples. At pressures below $2.7$ GPa, the data obtained in the Toroid-type pressure cell are consistent with the findings observed by the
piston/cylinder-type cell (Fig.~\ref{fig:s1}). The lower-temperature peak that is considered to be associated with the plaquette-singlet (PS) state can only be
detected completely by the Toroid-type pressure cell at $1.9$ GPa and $2.3$ GPa [Fig.~\ref{fig:s2}(b)], due to the fact that the lowest attainable temperature
of the cryostat used is $\approx 1.5$ K. At $2.1$, $2.4$, and $2.5$ GPa [Fig.~\ref{fig:s2}(a)] we can see up-turns at the lowest temperatures but the peak
is missed due to the restriction of the temperature range.

At pressures higher than $3$ GPa, a new transition was observed in the temperature range of $1.7-3.5$ K, which is considered to be related to an antiferromagnetic
(AF) transition. In one case, $P=4.0$ GPa in Fig.~\ref{fig:s2}(a), the peak associated with ordering is very clear, while in other cases the peak is rather
broad or shoulder-like, and, consequently, there is an uncertainty of order $0.2$ K in the transition temperatures graphed in Fig.~\ref{fig:phases}(a).
The small ordering peaks are expected for weakly coupled spin-isotropic two-dimensional antiferromagnets \cite{Sengupta2003}. We found that the transition
temperature of the AF phase shifts to higher temperature with increasing pressure, as also expected within the weakly-coupled SS layer description (as discussed
in the main paper).

Further compression leads to another previously not observed phase transition at $T \approx 8-9$ K between $P=4.1$ GPa and $4.9$ GPa; see Fig.~\ref{fig:s2}(b).
The previously known AF phase transition at higher temperature, above $100$ K \cite{Zayed2017,Loa2005}, was also found in our high-pressure heat capacity
studies with the Toroid-stype pressure cell, as we will discuss below.

\subsection{3. Extended pressure-temperature phase diagram}

We summarize our experimental results for the pressure measurements all the way up to 5 GPa in Fig.~\ref{fig:s3}(a), presenting an extension of the phase
diagram in the main paper, Fig.~\ref{fig:phases}(a),
with data above $4$ GPa added. Below $P \approx 4$ GPa, we have discussed three phases: the low-$T$ dimer-singlet (DS) state,
which is adiabatically connected to the high-temperature paramagnetic (PM) state, the PS state, and the AF state. The PS phase was expected in light of the
inelastic neutron scattering study by Zayed et al. \cite{Zayed2017}, who found a new excitation mode argued to show a PS state at $T=0.5$ K. However, the
phase boundaries had not been mapped out and recently the very existence of the PS phase in SrCu$_2$(BO$_3$)$_2$ was questioned \cite{Boos2019}.
In addition to finding what we argue is the PS phase, we identified the AF phase that had been expected based on the SS model but that was
previously never observed in the temperature and pressure regime found here;
starting at $P \approx 3$ GPa and extending to $P \approx 4$ GPa. The transition temperature
$T_{\rm AF}$ of the new AF phase varies from $\approx 2$ K to $\approx 3.5$ K increasing with $P$. This temperature scale of
the AF phase is reasonable within an SS description supplemented by weak inter-layer couplings, as discussed and illustrated with ED and QMC results
in the main paper. In contrast, it was previously believed that the AF phase starts only at $4$ GPa and has a transition temperature around $120$ K.
This temperature scale is unreasonably high within a description of weakly coupled SS layers, where one would expect the transition temperature
to be well below $J$ and $J'$, both of which should be of the order tens of K in the relevant pressure range. Thus, our study resolves a key
puzzle of the previously believed facts about SrCu$_2$(BO$_3$)$_2$---though this glaring mismatch was never emphasized as far as we are aware.

As shown in Fig.~\ref{fig:s3}, we also observe a phase transition at $T$ above $100$ K in out high-pressure measurements with the Toroid-type
pressure cell from pressures slightly above $4$ GPa up to the highest pressures studied, $P \approx 5$ GPa. As shown in Fig.~\ref{fig:s3}(a), at
$P = 4.15$ GPa, we observed this transition, into a phase that we will refer to as AF$_{\rm HT}$, at $T\approx 125$ K, consistent with the results reported
by Zayed et al.~\cite{Zayed2017}. At the same pressure, we further observe a second phase transition at $T \approx 8.2$ K. Such a transition was
not reported by Zayed et al.~\cite{Zayed2017}, who in their Fig.~S6 showed an AF order parameter increasing with decreasing $T$
down to $T \approx 12$ K. They also showed the presence of an AF Bragg peak at $T=4$ K. Thus, it appears likely that the new transition we
observe at $T \approx 8$ K (somewhat increasing with $P$) between $P = 4.2$ GPa and $4.9$ GPa is also AF in nature. We do not have any independent
evidence for antiferromagnetism in this state, which we therefore refer to as an unknown magnetic state (UM), but the low-temperature behavior
of $C/T$ in Fig.~\ref{fig:s3} at least indicates a gapless state. It could be an AF state with some minor difference---perhaps in the magnitude
of the order parameter---from the AF$_{\rm HT}$ state.

It appears most likely that both the AF$_{\rm HT}$ and UM phases arise from physics beyond the SS model. Given that a structural transition from tetragonal to
monoclinic has been long known within the pressure and temperature ranges of relevance here \cite{Loa2005,Zayed2014,Haravifard2014}, it is plausible that
the AF$_{\rm HT}$ and UM phases are both associated with the monoclinic crystal structure, in which the SS model does not provide an appropriate description.
Understanding the physics of this UM state and the AF$_{\rm HT}$--UM transition, in particular, deserves further investigations in the future. 

\null\vskip5mm
\subsection{4. Exact diagonalization of the Shastry-Sutherland model}

The temperature dependent heat capacity was calculated by standard numerical diagonalization \cite{Sandvik2010} of the SS Hamiltonian 
in all sectors of fixed total magnetization, $S^z=0,\pm 1,\ldots,\pm N/2$. The largest lattice on which we can fully diagonalize the SS 
Hamiltonian is $N=20$ spins; an often used tilted cluster on the square lattice \cite{Schulz1996}. The same lattice size was previously 
used for calculations of the uniform magnetic susceptibility in Ref.~\cite{Zayed2017}. The temperature dependent specific heat was computed 
directly as the Boltzmann-weighted expectation value $C=T^{-2}N^{-1}(\langle H^2\rangle-\langle H\rangle^2)$. 

\begin{figure}[t]
\begin{center}
\includegraphics[width=75mm]{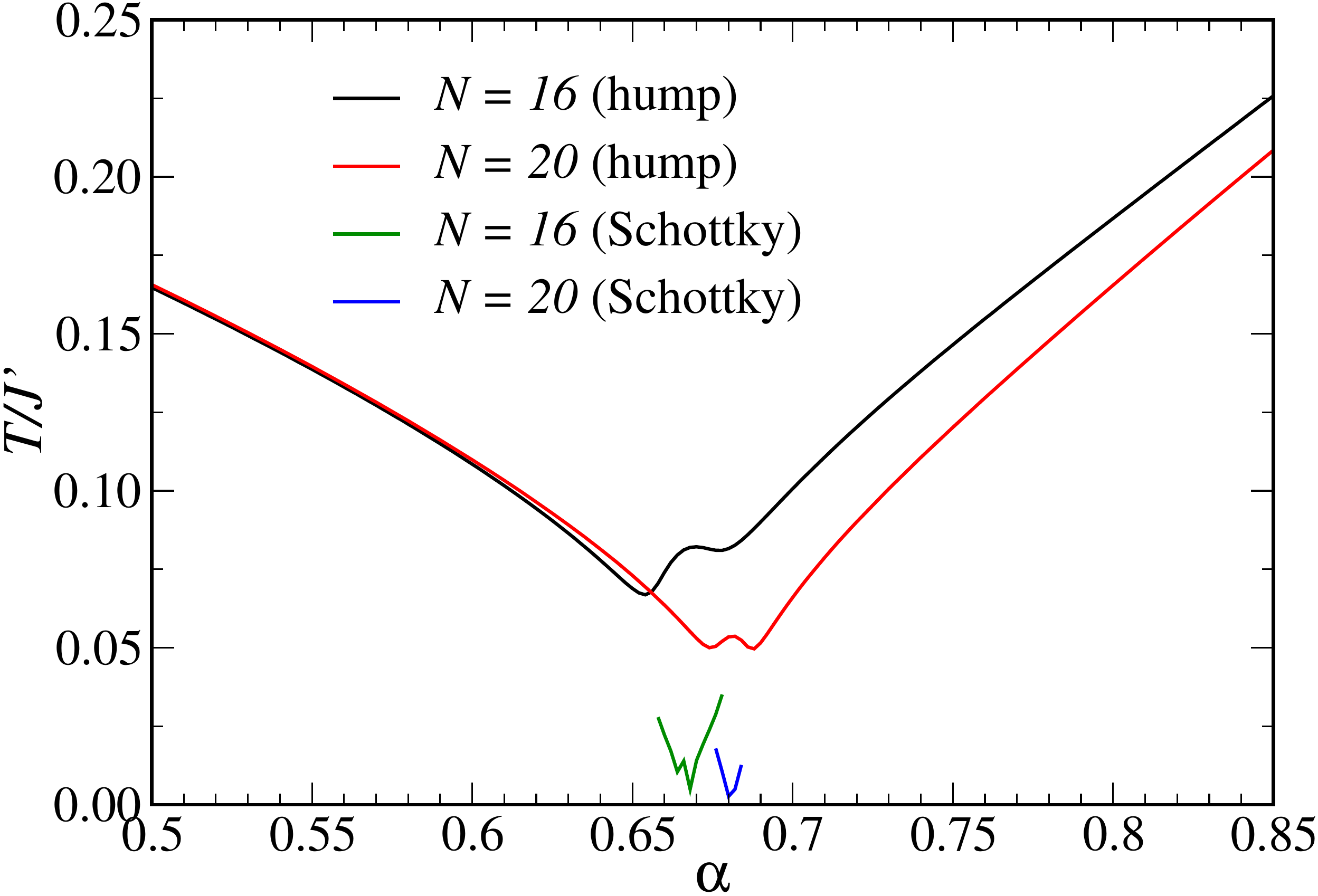}
\caption{The location of the hump in $C(T)/T$ graphed versus the coupling ration $\alpha=J/J'$ of the SS model on the $N=16$ and
$N=20$ clusters. The location of the Schottky anomaly in the $\alpha$ range where it is visible is also shown.}
\label{figs:th}
\end{center}
\end{figure}

Clearly thw small system cannot be expected to completely reflect the behavior in the thermodynamic limit, but in the large-gap DS phase the
remaining finite-size effects in $C(T)$ are small. In the PS phase, the peak corresponding to the phase transition can not yet be discerned.
Based on our work on the 2D CBJQ model, we know that much larger system sizes are required before this peak becomes prominent; see Sec.~5 below.
The main hump in $C(T)/T$, on which our comparisons between the SS model and the experiments are focused, should have much smaller
finite-size effects.

In Fig.~\ref{figs:th} we plot the hump temperature $T_{\rm h}$ versus the coupling ratio $\alpha=J/J'$ for system sizes $N=16$ ($4\times 4$ cluster)
and $N=20$. The $N=20$ data for $T_{\rm h}(\alpha)$ converted to the pressure dependent $T_{\rm h}(P)$ is shown in Fig.~\ref{fig:phases}(a) in the main
text. We used $P$-linear pressure dependent coupling constants $J(P)$ and $J'(P)$ as described in the caption of Fig.~\ref{fig:phases}. In Fig.~\ref{figs:th}
we can observe that the differences between $N=16$ and $N=20$ are small for $\alpha \alt 0.65$, i.e., when the system is well inside the DS phase.
As the PS phase is approached the size effects increase and persist inside the PS phase ($\alpha \approx 0.68 - 0.75$ \cite{Corboz2013})
and the AF phase. The main feature of a minimum in $T_h$ at $\alpha$ in the neighborhood of the DS--PS transition is present for both system sizes, however.
Finite-temperature properties eventually converge exponentially as a function of the system size, and most likely the hump temperature does not
move substantially away from the $N=20$ curve for larger system sizes. It would still be useful to study larger clusters in the future, e.g., with methods
such as those discussed in Refs.~\cite{Prelovsek2018,Li2018}.

\begin{figure}[t]
\begin{center}
\includegraphics[width=75mm]{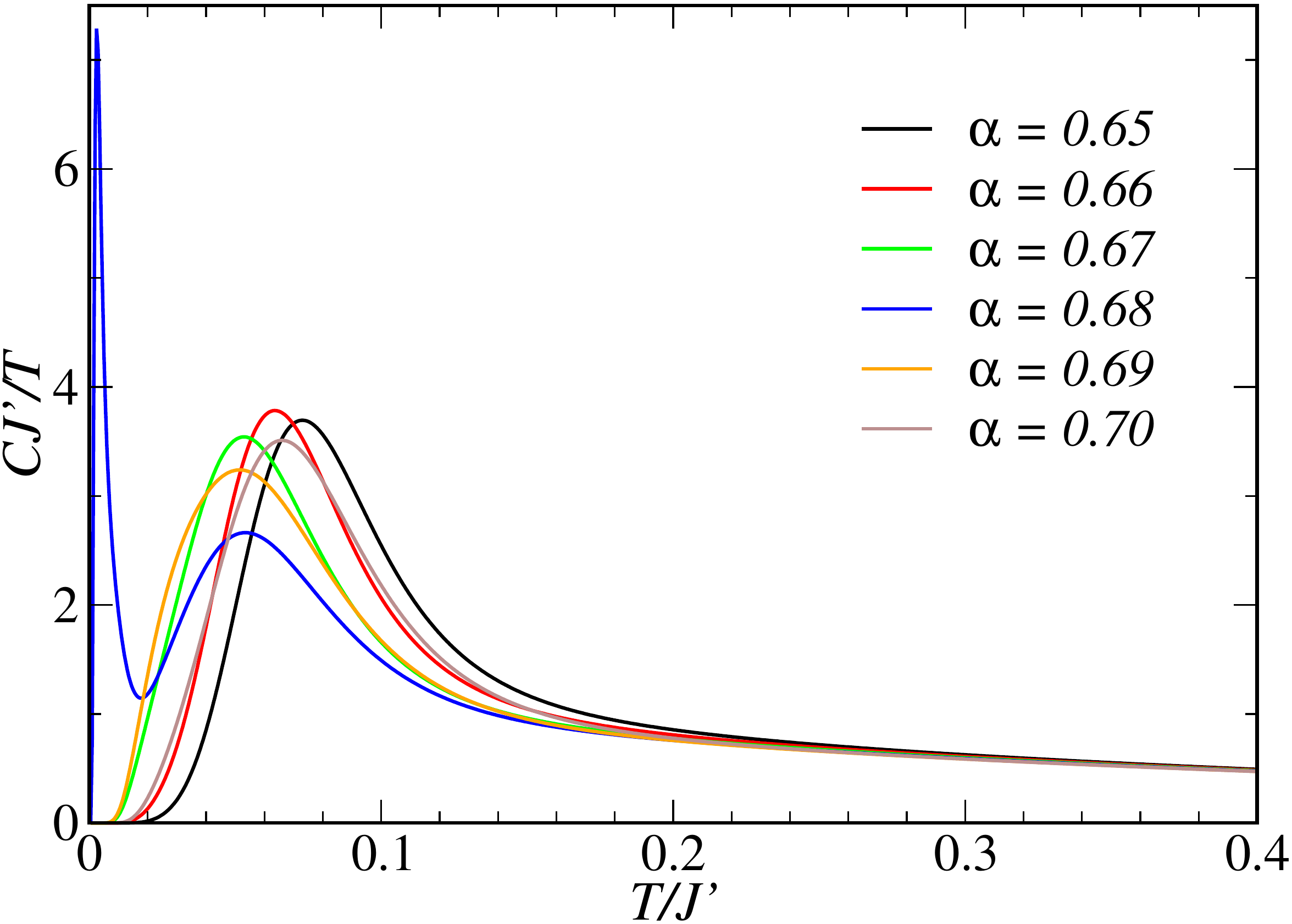}
\vskip-2mm
\caption{Temperature dependence of the heat capacity divided by the temperature of the $N=20$ SS cluster at several
values of the coupling ratio $\alpha$ of relevance in the comparisons with experimental data for SrCu$_2$(BO$_3$)$_2$.}
\label{figs:ctex}
\end{center}
\vskip-2mm
\end{figure}

In small clusters one can also observe a sharp low-temperature peak in $C/T$ that is related to the first-order DS--PS transition. This transition
is associated with a level crossing, and therefore a Schottky anomaly will be present in the heat capacity when the system is close to the phase
transition (when the two crossing levels are close to each other). The location of the Schottky peak is also indicated in Fig.~\ref{figs:th}.
Interestingly, for $N=20$, this peak temperature approaches zero at $\alpha \approx 0.68$, very close to the location of the DS--PS transition in
the thermodynamic limit \cite{Corboz2013}. Thus, already this small cluster can correctly reproduce the correct transition point. The use of level
crossing for accurate estimates of quantum phase transitions in 2D frustrated quantum spin models has recently been emphasized in Ref.~\cite{Wang2018}.

For completeness we also present $C(T)/T$ curves for several values of $\alpha$ in the $N=20$ cluster in Fig.~\ref{figs:ctex}. In addition to the
hump present for all $\alpha$ values shown, for $\alpha=0.68$ the prominent Schottky anomaly can also be seen at very low temperature. The other
cases are already sufficiently away from the phase transition for the two relevant levels to be far from each other and no anomaly can be observed.

\subsection{5. Checker-board $J$-$Q$ models}
 
In this section we provide additional information on the QMC simulations and finite-size scaling procedures underlying the phase diagrams
and $C/T$ curves of the 2D and 3D checker-board JQ (CBJQ) models in Fig.~\ref{fig:models}(a) in the main text.

As mentioned in the main text, the 3D CBJQ model is an extension of its 2D counterpart studied in Ref.~\cite{Zhao18}. The models are defined
using singlet projector operators,
\begin{equation}
P_{ij}=(1/4-{\bf S}_i \cdot {\bf S}_j),
\end{equation}
for nearest-neighbor $S=1/2$ spins. The 2D model is defined by the Hamiltonian
\begin{equation}
H_{\rm 2D} = - J\sum_{\langle i j \rangle}P_{ij}-Q \hskip-2mm \sum_{ijkl \in \Box^\prime}  \hskip-1mm (P_{ij} P_{kl}+P_{ik} P_{jl}),
\label{Eq:2DCBJQModel}
\end{equation}
where $J$ is equivalent to the standard Heisenberg interaction and $Q$ is the four-spin interaction present on every second plaquette (denoted
by $\Box^\prime$ above) in a staggered pattern as illustrated in Fig.~\ref{fig:models}(a). A small AF interlayer coupling $J_{\perp}$ is introduced
in the 3D model between identical 2D CBJQ systems with layer index $l=1,\ldots,L_z$,
\begin{equation}
H_{\rm 3D} = \sum_{l=1}^{L_z}H_{\rm 2D}(l) -J_\perp\sum_{\langle ij \rangle_{\perp}}P_{ij},
\label{Eq:3DCBJQModel}
\end{equation}
as also depicted in the schematic model illustration in Fig.~\ref{fig:models}(a) of the main text. We set $J+Q=1$ as the energy unit and define the
ratio $g=J/(J+Q)$ as our tuning parameter.

To simulate the models without approximations beyond statistical errors, we employ the SSE QMC method \cite{Sandvik2010}.
In 2D we study $L\times L$ square lattices with periodic boundary conditions, and in 3D we choose the size in the third direction as $L/2$, reflecting
the weak values of $J_\perp$ considered.

With the SSE method, the most direct way to compute the specific heat is from the fluctuations of the sampled expansion order $n$ \cite{Sandvik2010};
\begin{equation}
C=\frac{1}{N}(\langle n^{2}\rangle-\langle n\rangle^{2}-\langle n\rangle),
\label{spheat}
\end{equation}
where we normalize by the number of spins $N$. Alternatively, one can compute the internal energy $E(T)$ on a dense grid of $T$ points
and take the derivative $C(T)=dE/dT$ numerically. We have used both methods and find good agreement where they both work well---for low $T$
the derivative method is often preferrable as the statistical errors of the direct method increase rapidly as $T$ is lowered (more so than 
derivative estimators based on two or more temperatures).

\subsubsection{A. 2D CBJQ model}

The 2D CBJQ model was already discussed in detail in Ref.~\cite{Zhao18}; it exhibits a first-order quantum phase transition at $g \approx 0.18$
between the PS and AF ground states [note that a different definition of the tuning parameter was used, $g=J/Q$, and we have rescaled to the definition
$g=J/(J+Q)$ used in the present work]. Reflecting the unusual emergent O(4) symmetry found at this transition, the $T>0$ 2D Ising-type phase transition into
the PS state for $g < g_c$ has the form $T_c \propto 1/|\ln(g_c-g)|$ of the critical temperature, based on the analogy with a uniaxially deformed
O(4) model \cite{Irkhin1998}.

\begin{figure}[t]
\begin{center}
\includegraphics[width=75mm]{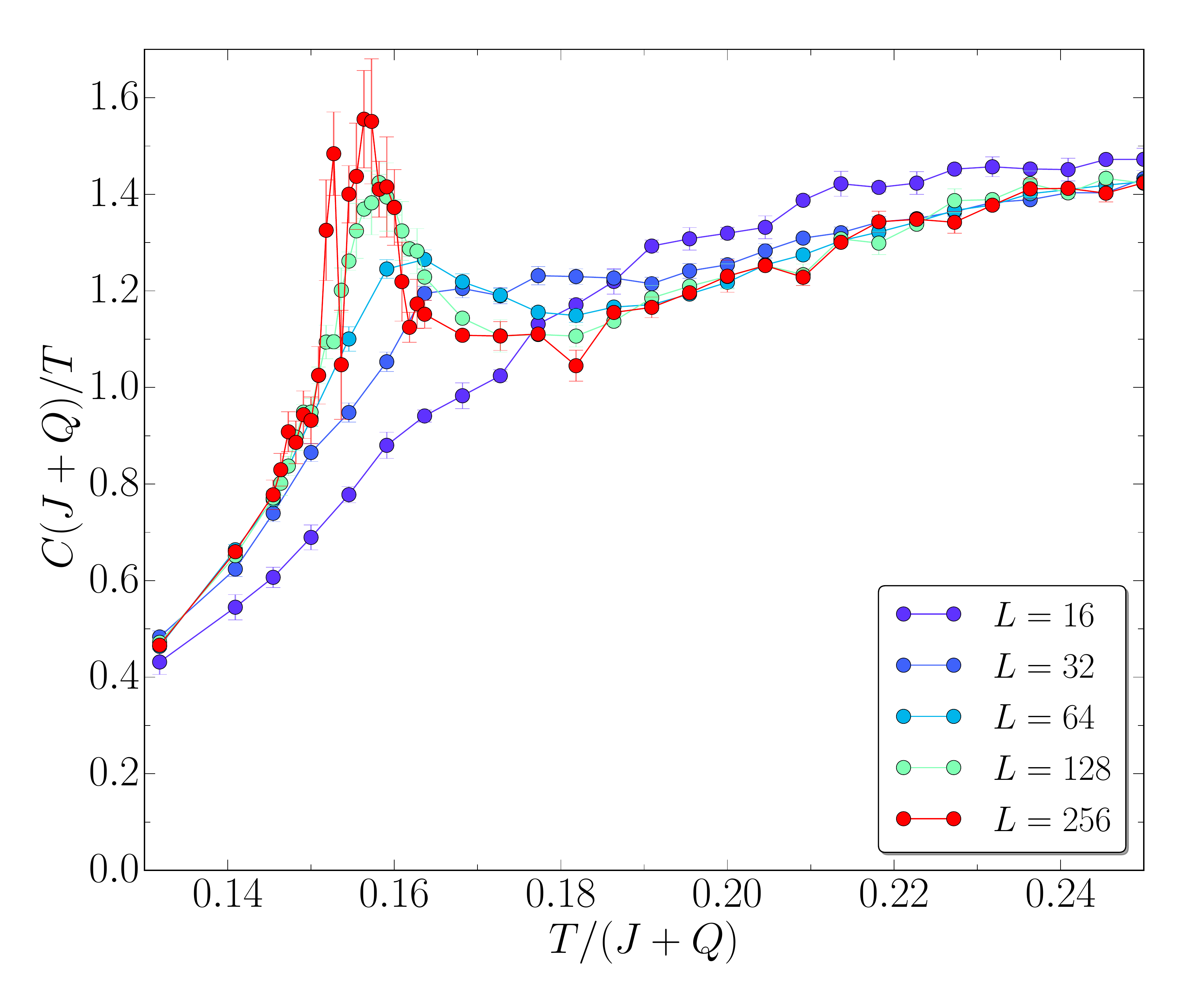}
\vskip-2mm
\caption{Heat capacity of the 2D CBJQ model with $g\approx 0.091$ ($J/Q=0.1$) in the neighborhood of its PS ordering transition. Results
  for several lattice sizes are shown in order to illustrate the size dependence of the peak associated with PS ordering.
  The transition is in the 2D Ising universality class, for
which the specific-heat exponent $\alpha=0$ and there is a logarithmic divergence of the peak value with the system size. The area under the
peak should converge to a finite value that vanishes as $g\to g_c$, $T_c \to 0$.}
\label{figs:ctl}
\end{center}
\vskip-2mm
\end{figure}

Here, in Fig.~\ref{figs:ctl} we present results for the heat capacity for a series of different lattice
sizes at $g \approx 0.09$ in order to systematically observe how the peak associated with PS ordering gradually emerges with increasing system size;
results for our largest system sizes were shown in Fig.~\ref{fig:models}(b) in the main text. No ordering peak can be discerned at all for $L=16$.
Thus, the absence of ordering peak in the ED results for the $N=20$ SS model in the PS range of $\alpha$
values on much smaller lattices [Fig.~\ref{figs:ctex}] is not surprising.

Note that the 2D model does not exhibit any AF order at $T>0$, only exactly at $T=0$, as a consequence of the Mermin-Wagner theorem according
to which a continuous symmery, here O(3) spin-rotation symmetry, cannot be broken at $T>0$ in two dimensions. We therefore use the 2D model only
to elucidate the PS state and transition in SrCu$_2$(BO$_3$)$_2$.

\subsubsection{B. 3D CBJQ model}

In the 3D case the CBJQ model can undergo AF ordering also at $T>0$. If the relative intra-layer coupling $J_\perp/(Q+J)$ is small, we expect
a separation in temperature between a hump in $C(T)/T$ and the peak associated with the phase transition, as was previously studied with SSE
QMC simulations in the Heisenberg case ($Q=0$) \cite{Sengupta2003}. We will demonstrate this seperation of temperature scales here. In addition,
we investigate the sensitivity of the location of the PM--PS and PM--AF phase boundaries, as well as the direct PS--AF boundary, to variations
on $J_\perp$. We have performed simulations for $J_{\perp}=0.1$ and $0.01$ (with $J+Q=1$).

To capture the finite-temperature phase transitions from the PM phase into the PS and AF phases, we calculate the Binder cumulants
of the respective order parameters, defined as
\begin{align}
U_z &= \frac{5}{2} \left(1 - \frac{\langle m_{z}^4\rangle}{3\langle m_{z}^2 \rangle^2}\right)\label{cumdef1}\\
U_p &= \frac{3}{2} \left(1 - \frac{\langle m_{p}^4\rangle}{3\langle m_{p}^2 \rangle^2}\right),
\label{eq:cumdef2}
\end{align}
where $m_{z}$ and $m_{p}$ are the order parameters for AF and PS phases. The AF order parameter $m_z$ is taken as the $z$-component of the O(3)
staggered magnetization vector,
\begin{equation}
m_{z} = \frac{1}{N}\sum_{i} (-1)^{x_i+y_i+z_i} S^z(i),
\label{msdef} 
\end{equation}
where $S^z(i)$ is the spin at site $i$ with coordinates $(x_i,y_i,z_i)$ on the 3D cubic lattice. As for the plaquette order parameter, we first definte
its $l$:th layer value as
\begin{equation}
 m_p(l) = \frac{2}{L^2}\sum_{i \in \Box^\prime} \phi(i) \Pi^z(i),
\label{mpdef} 
\end{equation}
where the sum is over the $Q$-plaquettes $\Box^\prime$ and $\phi(i)=\pm 1$ for even and odd rows of plaquettes. The plaquette quantity
$\Pi^z(i)$ is defined as
\begin{equation}
\Pi^z(i)=S^z(i) S^z(i + \hat{x}) S^z(i + \hat{y})S^z(i + \hat{x} + \hat{y}),
\label{plaquettez}
\end{equation}
where the site $i$ stands for the low-left corner site in a given $Q$ plaquette. The full 3D order parameter used in the Binder cumulant in
Eq.~\eqref{eq:cumdef2} is defined as the average of $m_p(l)$ over the layers, 
\begin{equation}
m_p=\frac{1}{L_{z}}\sum\limits_{l}^{L_{z}} m_{p}(l),
\label{mpdef3d} 
\end{equation}
with $L_z=L/2$. Note that the 3D PS order parameter defined in Eq.~(\ref{mpdef3d}) corresponds to in-phase ordering of the plaquettes within
the different layers, which is what we find in this version of the 3D CBJQ model. Out-of-phase ordering could be achieved by modifying the inter-layer
coupling.

\begin{figure}
\includegraphics[width=72mm]{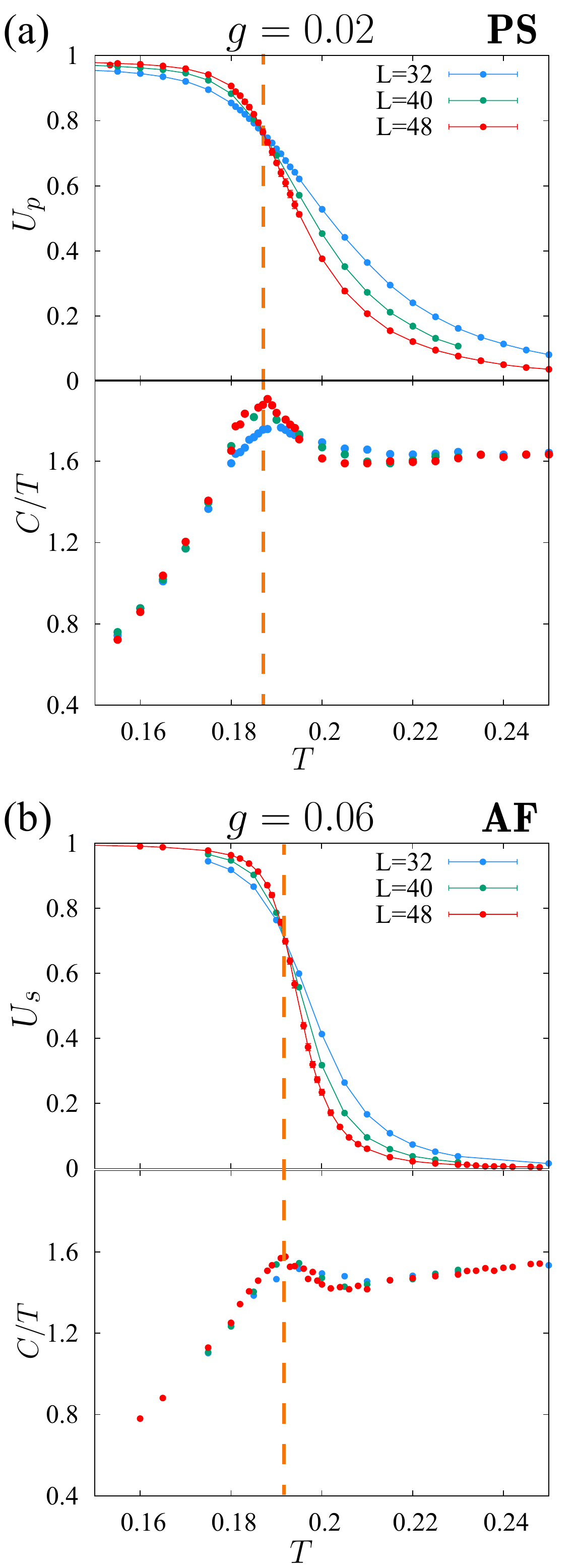}
\caption{Temperature dependence of the plaquette and spin Binder cumulants,
$U_p$ and $U_s$ (upper panels), and the heat capacity $C/T$ (lower panels) of the 3D CBJQ model
with $J_{\perp}=0.1$, calculated on lattice sizes $L=32,40$, and $48$. The in-plane coupling ratio is $g=0.02$ in (a) and $0.06$ in (b), corresponding,
respectively, to the ordered PS and AF phases at low temperatures. The orange dashed lines mark the common location of the crossing points of the
Binder cumulants and the peak in $C/T$, i.e., the transition temperature $T_c$. The humps located above $T_c$ are seen more clearly on the wider
temperature scale used in Fig.~\ref{fig:models}(c) in the main paper.}
\label{fig:supfig2}
\end{figure}

\begin{figure}
\includegraphics[width=70.5mm]{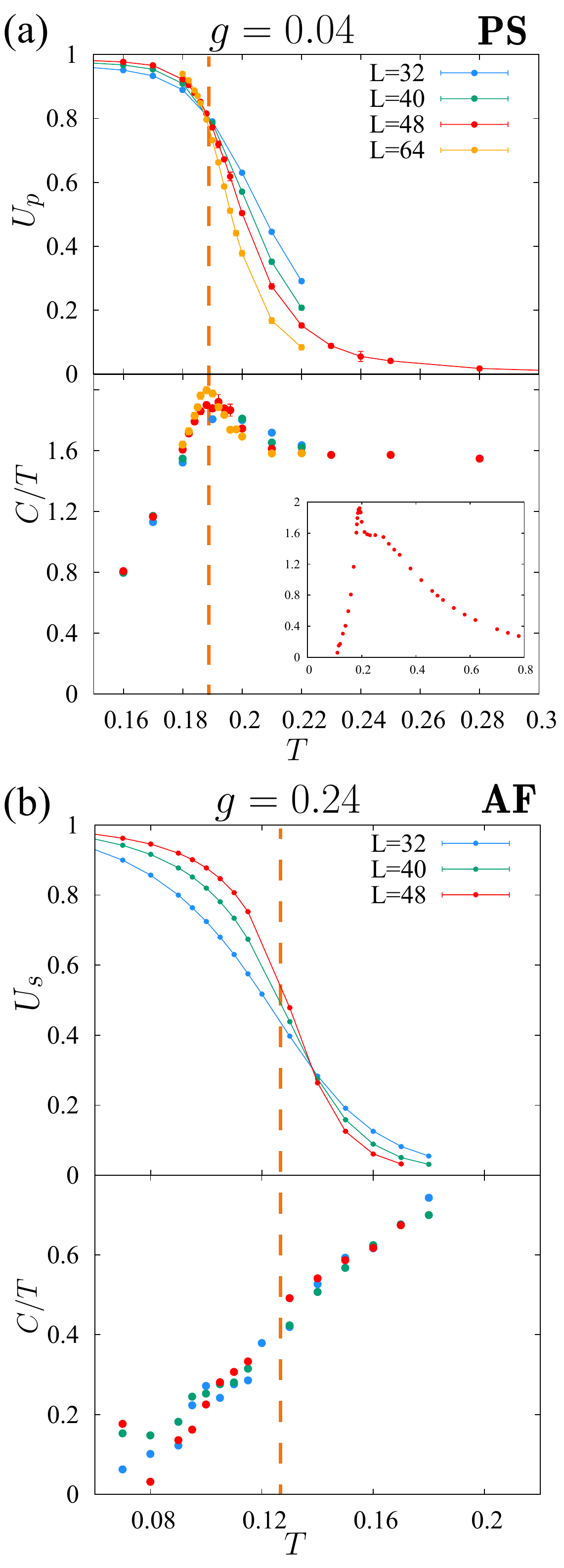}
\caption{Results analogous to those in Fig.~\ref{fig:supfig2} for a much weaker inter-layer coupling; $J_\perp=0.01$. Here the coupling ratio
$g=0.04$ in (a), corresponding to the PS phase at low $T$, and the system sizes are $L=32,40,48$, and $64$. The inset shows the $L=48$ data
on a wider $T$ scale. In (b), at $g=0.24$ the system is in the AF phase at low $T$ and the lattice sizes are $L=32,40$, and $48$. At the AF transition
the ordering peak is very small and not clearly discernible where it should appear at the orange line. Here the hump at higher $T$ is not seen
because of the lack of data, but it should be located at $T$ above $0.2$ as in Fig.~\ref{fig:models}(c) in the main paper.}
\label{fig:supfig3}
\end{figure}

The phase transitions are located by the common method of Binder cumulant crossings; scanning over $T$ or $g$, the cumulants for two different system
sizes cross at some point close to the phase transition, where in the thermodynamic limit the cumulant for a given order parameter exhibits a step
function, jumping from $0$ in the phase with no order of the type considered to $1$ when there is such order. The crossing points for different pairs of
system sizes will flow to the location of the step as the system size is increased.

At a conventional first-order transition, the cumulant develops a negative divergent peak at a location that also flows toward the transition point.
No negative peaks were found at the $T=0$ PS--AF transition in the 2D CBJQ model \cite{Zhao18}, even though other first-order signatures are clearly
visible. This anomalous behavior, in combination with other considerations, led to the conclusion of a first-order transition with emergent O(4) symmetry.
In a forthcoming paper we will investigate the fate of the emergent symmetry in the 3D CBJQ model with weakly coupled layers \cite{Sun2019}.
Here we focus on the phase diagram and the behavior of the specific heat, complementing the results that allowed us to connect to
the experiments on SrCu$_2$(BO$_3$)$_2$ in the main paper.

\begin{figure}[t]
\includegraphics[width=80mm]{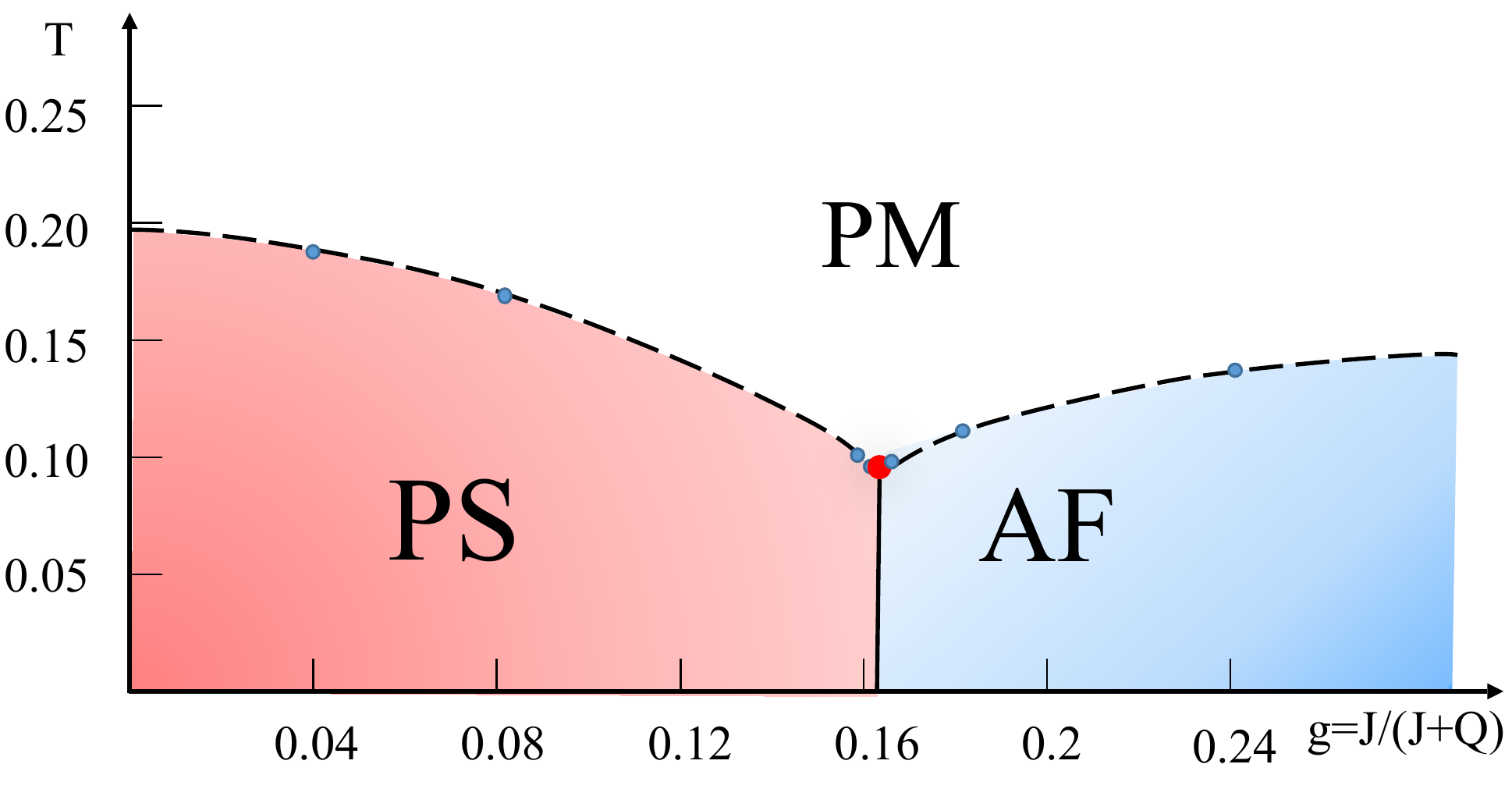}
\caption{Phase diagram of the 3D CBJQ model with inter-layer coupling $J_{\perp}=0.01$. The PM, PS, and AF phases are the same as those in the
Fig.~\ref{fig:models}(c) of the main text at $J_{\perp}=0.1$. The dashed lines nark continuous phase transition and the solid line indicates
a first order transtion. The phase boundaries were obtained by interpolating among transition points extracted from Binder cumulant crossings,
such as those in Fig.~\ref{fig:supfig3} and additional scans vs $g$ at fied $T$.}
\label{fig:supfig4}
\end{figure}

At each fixed $g$, we perform simulations scanning the $T$-axis for various system sizes. The finite-size analysis can be used to determine the critical
temperature $T_c$ and the divergence/singularities of thermodynamical quantities. Figures \ref{fig:supfig2} and \ref{fig:supfig3} show representative
results for $J_{\perp}=0.1$ and $0.01$, respectively. In Fig.~\ref{fig:supfig2}(a), at $g=0.02$ the system is inside the PS phase at low temperature. The
PM--PS phase transition is manifested as the crossings of $U_p$ curves for different system sizes. With three different system sizes, $L=32,40,48$,
the crossing can be determined at $T_c \approx 0.187$, as denoted by the orange dashed line. At the same temperature, $C/T$ develops a weak divergence,
as seen in the lower panel of Fig.~\ref{fig:supfig2}(a). This behavior is expected at a 3D Ising critical point, where the specific-heat
exponent $\alpha$ is close to $0$ but positive. The hump above the $C/T$ peak is more clearly observable on the wider $T$ range shown in
Fig.~\ref{fig:models}(c). The hump exhibits only a weak size dependence, reflecting the short correlation length at these temperatures. The same
kind of peak-hump structure is also observed in the 2D case [Fig.~\ref{fig:models}(b) in the main paper and Fig.~\ref{figs:ctl} above] and in the
experiments on SrCu$_2$(BO$_3$)$_2$ [Fig.~\ref{fig:phases}(c)], suggesting that these features are largely developing due to correlations and interactions
within the 2D layers. The 3D couplings still play an important role quantitatively, especially in the significant shrinking of the PS phase relative
to the purely 2D case---the same mechanism reduces the critical coupling ratio $\alpha$ of the PS--AF transition in the SS model when the $J_\perp$
interactions are tunrned on, as discussed in the context of fitting to experimental SrCu$_2$(BO$_3$)$_2$ data in the main paper.

Increasing the value of $g$ to $0.06$, in the AF regime, we can see in Fig.~\ref{fig:supfig2}(b) that the $U_s$ curves cross at
$T_c \sim 0.191$. At the same temperature, $C(T)/T$ also develops a peak, corresponding to a continuous transition into the AF phase. In this case
we do not expect a divergent peak as $L$ increases, only a cusp singularity corresponding to the small negative value of the exponent $\alpha$
in the 3D $O(3)$ universality class. Indeed, the peak shape does not change significantly with the system size in this case.
The broad hump slightly above the peak, signifying the onset of 2D magnetic fluctuations, is also observed. In the AF phase the 3D couplings
clearly play a crucial role in determining the shape of the $C/T$ curve, as the ordering transition is completely absent for an isolated
2D layer.

In Fig.~\ref{fig:supfig3} we show results similar to those above for $J_{\perp}=0.01$. Figure \ref{fig:supfig3}(a) corresponds to the PM--PS transition
at $g=0.04$, where the crossings of $U_p$ curves give the transition temperature $T_c \approx 0.187$; almost the same as in the $J_{\perp}=0.1$ case.
This confirms again the minimal impact of a weak inter-plane coupling in the PS phase relatively far away from the 2D quantum-critical point.
The AF ordering temperature, analyzed in Fig.~\ref{fig:supfig3}(a), is much more affected, being reduced from $T_c \approx 0.19$ to $\approx 0.14$
when $J_\perp$ is decreased from $0.1$ to $0.01$. The still very high critical temperature in units of $J_\perp$ reflects the expected form
$T_c \propto 1/|\ln(g-g_c)|$ \cite{Irkhin1998}.

Finally, in Fig.~\ref{fig:supfig4} we present the phase diagram of the 3D model at $J_\perp=0.01$, complementing the phase diagram at $J_\perp=0.1$ in
Fig.~\ref{fig:models}(c) of the main paper. The phase boundaries were drawn based on several scans of the type shown in Fig.~\ref{fig:supfig3},
and additional scans at fixed $T$ and carying $g$. The quantum phase transition between the PS and AF phases takes place at $g=0.162$,
roughly $10\%$ smaller than the $g_c$ value in the $J_{\perp}=0$ case, demonstrating that even a very weak inter-layer coupling can noticably affect the location
of the quantum phase transition, as we have argued in the case of the SS model in the main paper based on the results for the 3D CBJQ model
presented here.

The $T>0$ bicritical point at which the first-order AF--PS transition terminates is at $(g,T) \approx (0.162,0.11)$, marked with the
red circle in the phase diagram Fig.~\ref{fig:supfig4}.
This can be compared with the point $(g,T) \approx (0.041,0.168)$ for $J_{\perp}=0.1$. The bicritical point should fall
within the symmetry classification discussed in the context of classical models with $O(N_1)$ and $O(N_2)$ transitions, where here $N_1=1$ (the PS order
parameter) and $N_2=3$ (the AF order parameter), but we have not yet confirmed the scenario proposed for these particular values of $N_1$ and $N_2$
\cite{Eichhorn2013}.
  
\end{document}